\documentclass[baaa]{baaa}

\usepackage[pdftex]{hyperref}
\usepackage{subfigure}
\usepackage{natbib}
\usepackage{helvet,soul}
\usepackage[font=small]{caption}
\usepackage{paralist}

\begin{document}


\journalvol{61A}
\journalyear{2019}
\journaleditors{R. Gamen, N. Padilla, C. Parisi, F. Iglesias \& M. Sgr\'o}


\contriblanguage{1}


\contribtype{2}

\thematicarea{5}

\title{Sidelobe analysis for the Atacama Cosmology Telescope:}
\subtitle{a novel method for importing models in GRASP}


\titlerunning{Sidelobe analysis for the Atacama Cosmology Telescope}


\author{R. Puddu\inst{1}, N.F. Cothard\inst{3},  P.A. Gallardo\inst{2}, R. D\"unner\inst{1}, P. Flux\'a\inst{1}
for the ACT Collaboration}
\authorrunning{Puddu et al.}

\contact{rpuddu@aiuc.puc.cl}

\institute{
Instituto de Astrof\'isica y Centro de Astro Ingenier\'ia, Pontificia Universidad Cat\'olica, Santiago de Chile \and
Department of Physics, Cornell University, Ithaca, NY 14853, USA \and
Department of Applied and Engineering Physics, Cornell University, Ithaca, NY 14853, USA
}


\resumen{
Los telescopios para observar el fondo c\'osmico de microondas suelen tener pantallas y \textit{baffles}, con el fin de reducir la emisi\'on del suelo. Estas estructuras pueden introducir l\'obulos laterales no deseados. En este manuscripto introduciremos una herramienta en GRASP para obtener un modelo electromagn\'etico de objectos de gran tama\~no y estructura compleja, como los telescopios, y predecir la forma as\'i como la posici\'on de los l\'obulos laterales.}

\abstract{
Telescopes for observing the Cosmic Microwave Background (CMB) usually have shields and baffle structures in order to reduce
the pickup from the ground. These structures may introduce
unwanted sidelobes. We present a method to measure and model baffling structures of large aperture telescope optics to predict the sidelobe pattern.}



\maketitle

\section{The Atacama Cosmology Telescope}\label{ACT_intro}

The Atacama Cosmology Telescope (ACT) is a 6 m telescope located in northern Chile, at 5200 m elevation. The high altitude and low water vapor content in the atmosphere provide excellent conditions for observing the sky in mm wavelengths. The tropical latitude allows access to more than half of the sky. The design is an off-axis Gregorian. The large aperture is required to achieve arcminute resolution, which allows to study the Sunyaev Zel'dovich effect in galaxies clusters, weak lensing and give constraints on neutrino mass [\cite{sz, neutrino, lensing, lensing2}], as well as the CMB power spectrum up to high multipoles [\cite{Louis}]. The short focal length, 5.2 m, makes the telescope compact enough for fast scanning. To minimize
ground pick-up during scanning, the telescope has two ground
screens. A large, stationary outer ground screen surrounds the
telescope. A second, inner comoving screen connects the edges of the primary reflector to the secondary reflector and
moves with the telescope during scanning. A climate-controlled
receiver cabin is situated underneath the primary and secondary
reflectors [\cite{thornton, optical_design}].

Spillover at the Lyot stop can load the detectors with radiation emitted
from warm, nearby structures. Theoretical estimates based on the optical design show that there is
2\% spillover past the secondary reflector [\cite{swetz}]. To reduce this
loading, the primary reflector is surrounded by a 0.75 m radial baffle that
reduces the primary spillover to less than 0.2\%, by directing those rays to the sky. However, the spillover
on the secondary reflector occurs at larger angles. 
Measurements suggested that there is as much as 2\% to 3\% power that does not get reflected to the sky. As a result, the secondary baffling
was redesigned to ensure that the majority of the spillover was redirected to the sky [\cite{swetz}].																

\section{Sidelobes characterization}

Diffracted and reflected power may reach the sky in arbitrary angles, forming far sidelobes. If these sidelobes get illuminated by strong sources, like the sun or the moon, they may produce spurious features in maps of the sky. For large optical systems the sidelobe response can be challenging to model and characterize [\cite{Lockman,Page,Tauber,Carlstrom}]. Nevertheless, an accurate model of this systematic effect is crucial to obtain clean maps and accurate astrophysical results [\cite{Barnes}]. In order to characterize the sidelobes of ACT, we have performed simulations with the GRASP (General Reflector Antenna Software Package \footnote{\url{https://www.ticra.com/software/grasp/}}) physical optics software. Propagation of electromagnetic waves, as well as their reflection and refraction, is ruled by Maxwell's equations and induced currents play an important role in them. GRASP is able to compute surface currents induced by incoming radiation upon reflectors. These currents will cause the reflector to re-emit radiative power and thus can illuminate further reflectors. In this way it is possible to build any desired optical chain. The last reflector will radiate the complex fields, which will be evaluated on a grid, typically in the far field. Knowing the complex fields may give important insights on the polarization state and coherence of the signal, which are hard to obtain with other ray-trace similar software packages.

A model of the telescope is required in GRASP. This consists, in our case, in a series of (according to software notation) ``geometrical objects'' called ``reflectors''. Each element of the inner and outer shield, as well as the mirrors and the secondary baffling is treated as a reflector. Drawing the ACT model in GRASP is not trivial since every reflector must be defined with a coordinate system, a surface and a rim. This, unfortunately, cannot be done in a quick and intuitive way as it would be in Computer Aided Design (CAD) software. Information about the geometry of the model is handled with a file with specific GRASP syntax. Furthermore, while mechanical specifications are available for most of telescope parts, that is not the case for the secondary baffling.
For this reason we used photogrammetry to get a \emph{points cloud} recontruction of the telescope, which was used to build the model within a CAD software (SolidWorks). Later, we imported it into GRASP by means of a code able to read the CAD file and writing the corresponding file for the GRASP project.

\section{Photogrammetry}
In a photogrammetric system, a series of stickers, \emph{photogrammetry targets}, are photographated by a camera and processed by a suitable software. We have used a Nikon 600D camera and the Photoscan software.
The software combines a large amount of photos, performs triangulations for all recognized markers, and generates their 3D positions. The final result is a point cloud which reproduces the telescope shape.

\begin{figure}
\centering
\includegraphics[width=0.35\textwidth]{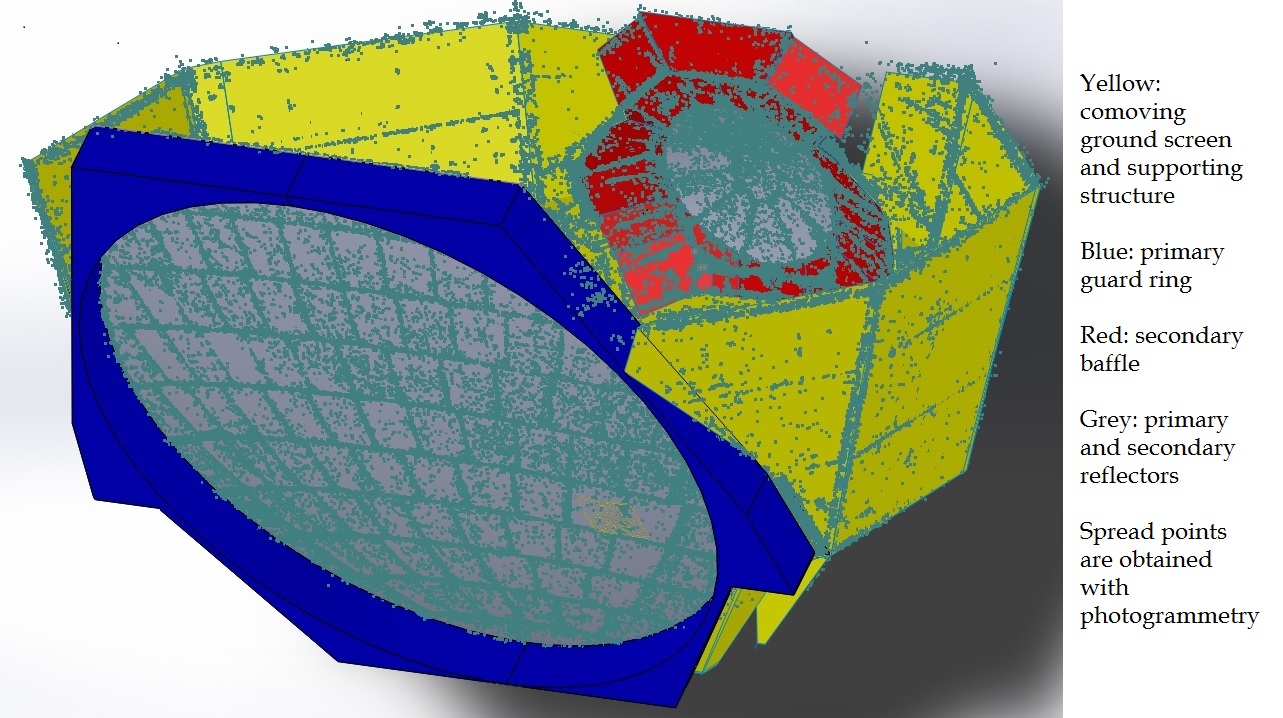}
\caption{Point cloud obtained with photogrammetry Photoscan software, overlapped with the ACT telescope CAD model. Most relevant parts in the GRASP simulations are the primary guard ring (blue) and the secondary baffle (red).}
\label{fig:CAD_model}
\end{figure}

We use the measured positions of the targets (the point cloud) to create a best fit model that represents the shape of the upper structure's planar faces. The planar faces fit the point cloud (see Fig. \ref{fig:CAD_model}). The primary and secondary reflectors have been drawn following the geometrical specifications of the telescope and not using the point cloud [\cite{optical_design}].

\section{ACT in GRASP}
\label{grasp}
GRASP is a software able to perform electromagnetic simulations of antennas. Its main purpose is to study the behavior of relatively simple systems composed by one or two reflectors, and a feeding device with potentially some more parts such as plates, struts, stiffeners, support rings and so on, i.e. a series of components that are useful for engineering purposes in several antenna applications. Instead, with this study we want to apply the computing potential of the software in order to investigate the systematics of the ACT, such as sidelobes. These are usually generated by interactions between optical elements such as reflections off baffling structures, scattering off optical elements, spurious reflections off support structures and panel gap diffraction. In ACT, most sidelobe contribution come from the baffle around the secondary mirror and the guard ring around the primary mirror [\cite{pato}]. The geometrical definition of these structures is a tricky obstacle, since the analysis of secondary effects like telescopes sidelobes is not the main purpose of the software.

We use the CAD model of ACT obtained with the point cloud to obtain a correspondent model in GRASP. This has been accomplished by means of a dedicated code able to perform such a conversion. A powerful Python library, FreeCAD \footnote{\url{https://www.freecadweb.org/}}, extracts the geometrical information contained in the CAD model. It reads the file extracting the model geometry in a hierarchical architecture, i.e. the solid body is organized in faces, each of them in lines, each line in points. All elements in the CAD file are referred to a unique coordinate systems, whilst in GRASP this is not the optimal scenario: this would both increase the computation time for currents and fields, and make the model hard to handle or modify. Thus, we assign a local coordinate system to each face of the CAD model. Local coordinate systems are referred to the global one via a suitable Euler rotation $(\phi, \theta, \psi)$ with axes $zyz$ (as defined in GRASP). In order to find $\phi$ and $\theta$ ($\psi$ is set to zero, as two angles are enough to define a $xy$ plane containing all the points of the face) we proceed as follows, for each face of the body:
 \begin{compactitem}
 \item any three consecutive points are taken to define two non-parallel vectors. Their cross product will define the normal vector $\mathbf{n}=(n_x, n_y, n_z)$ to the plane containing the face. The normal vector will be parallel to the $z$ axis of the local coordinate system, which will pass through the center of mass of the face.
 \item the center of mass of the face is set as the origin of the local coordinate system.
 \item $\phi$ is retrieved as $\arctan \frac{n_y}{n_x}$.
 \item $\theta$ is retrieved as $\arctan \frac{n_z}{\sqrt{n_x^2+n_y^2}}$.
 \item each point is translated by the vector of the center of mass and then rotated with the Euler angles $\phi$ and $\theta$. This ensure that all the points will lay in the $xy$ plane of the local coordinate system, i.e. $z_{point}=0$.
 \item the local system allows to define the surface easily, being the origin $(0,0)$ and the $z$ axis unit vector the constant parameters.
  \end{compactitem}

\begin{figure}
  \centering
  \includegraphics[width=0.35\textwidth]{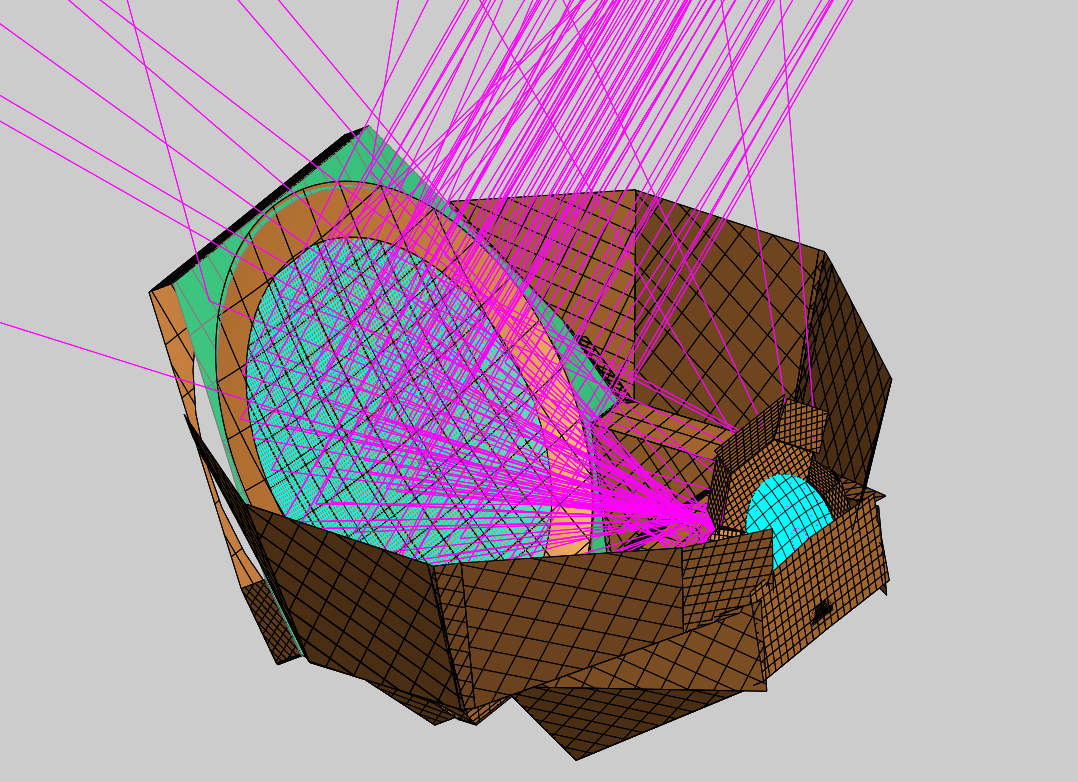}
  \caption{The same model as in Fig. \ref{fig:CAD_model}, imported in GRASP with our Python code.}
  \label{fig:CAD_model_Grasp}
  \end{figure}

  In this way all the geometrical elements are defined in a convenient way, improving the performance of the software and making the project more readable, allowing straightforward modifications, when necessary (e.g a rim point, or the tilt of a plane with respect to the global system coordinates).
  The importance of this code mostly relies in allowing to perform optical simulations with GRASP physical optics, including elements not previously available and thus improving the accuracy. In particular elements such the ones responsible for the sidelobes were tricky to insert in the model because of the non-trivial definition in GRASP.
  The model as it has been imported into GRASP is shown in Figure \ref{fig:CAD_model_Grasp}.

  \subsection{GRASP analysis}
  The secondary baffling, together with the large structure around the primary mirror and the outer part the primary mirror (the guardring, highlighted in green in Figure \ref{fig:CAD_model_Grasp}) are expected to be the most responsible of sidelobes (see Sec. \ref{ACT_intro}).
  The analysis has been performed in GRASP by illuminating the secondary baffle with a feed placed at the entrance window of the camera. The beam of the camera has been measured [\cite{pato}] \textit{in situ} and it is reproduced in GRASP. The secondary baffling in turn illuminates the guard ring and finally the fields are collected on the output, far field grid. A remarkable match arises comparing the simulated maps in GRASP with the observed ones and will be released in a future publication.

  \begin{figure}[h!]
  \centering
  \includegraphics[width=0.35\textwidth]{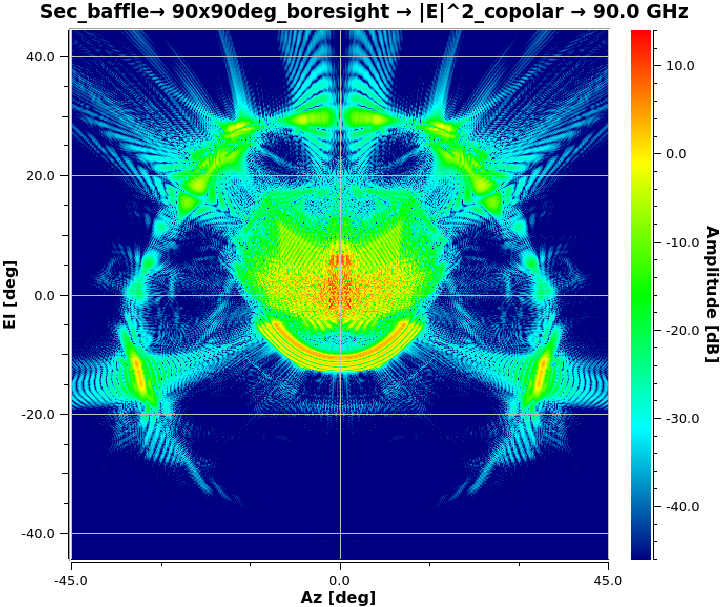}
  \includegraphics[width=0.35\textwidth]{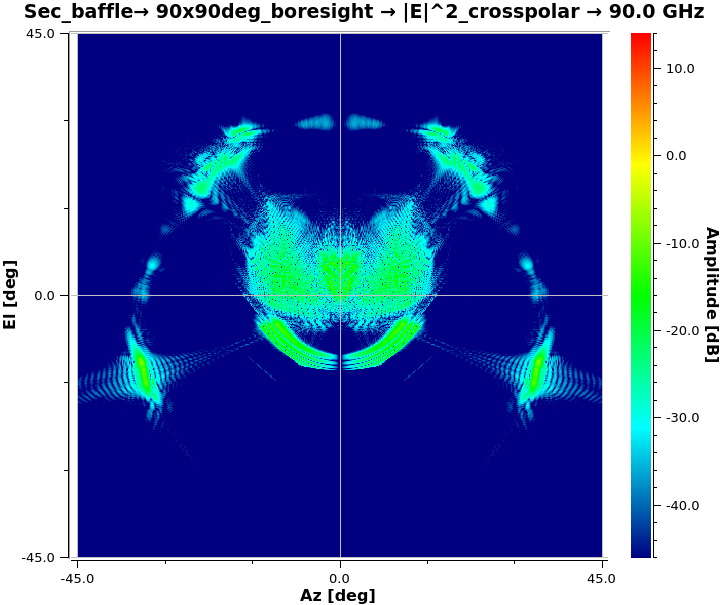}
  \caption{Far field of the telescope, as observed illuminating the baffle around the secondary mirror with subsequent scattering off the primary mirror and its guard ring structure. The angular coordinates azimuth and elevation are referred to the boresight, whilst the dB scale is referred to the injecting power. }
  \label{fig:power}
  \end{figure}

  \section{Conclusions and future steps}
  The utility developed to import CADs in GRASP makes possible to perform extensive simulations of sidelobes, thus allowing a complete characterization of the telescope. We were able to find the origin of the sidelobes, and this will let us decide the best strategy, both for scanning the sky and determining the panels that have the most trouble. In wintertime, when the humidity is very low and the weather conditions are stable, we are planning to perform extensive measurements involving all the panels, as well as to make sidelobes maps with the eccosorb on the panels to verify that contributors to the sidelobe structure have been appropriately identified and mitigated.

  \begin{acknowledgement}
  This work was supported by the CONYCIT award Anillo ACT-1147, BASAL CATA PFB-06, Anillo ACT-86, FONDEQUIP AIC-57 and QUIMAL 160009. We also acknowledge the U.S. National Science Foundation through awards AST-0408698 for the ACT project and PHY-0355328. We gratefully thank the supporting TICRA staff for their technical suggestions.
  \end{acknowledgement}


  \bibliographystyle{baaa}
  \small
  \bibliography{biblio}

\begin{thebibliography}{}
\catcode`\~=\active \def~{\penalty10000 \ }

\bibitem[\protect\citeauthoryear{{Barnes}, {Hill}, {Hinshaw}, {Page},
  {Bennett}, {Halpern}, {Jarosik}, {Kogut}, {Limon}, {Meyer}, {Tucker},
  {Wollack} \& {Wright}}{{Barnes} et al.}{2003}]{Barnes}
{Barnes} C.,  et al., 2003, \apjs, 148, 51

\bibitem[\protect\citeauthoryear{{Carlstrom}, {Ade}, {Aird}, {Benson}, {Bleem},
  {Busetti}, {Chang}, {Chauvin}, {Cho}, {Crawford} \& {Crites}}{{Carlstrom} et
  al.}{2011}]{Carlstrom}
{Carlstrom} J.~E.,  et al., 2011, Publications of the Astronomical Society of
  the Pacific, 123, 568

\bibitem[\protect\citeauthoryear{{Fowler}, {Niemack}, {Dicker}, {Aboobaker},
  {Ade}, {Battistelli}, {Devlin} \& {Fisher}}{{Fowler} et
  al.}{2007}]{optical_design}
{Fowler} J.~W.,  et al., 2007, \ao, 46, 3444

\bibitem[\protect\citeauthoryear{{Gallardo}, {Cothard}, {Puddu}, {D{\"u}nner},
  {Koopman}, {Niemack}, {Simon} \& {Wollack}}{{Gallardo} et al.}{2018}]{pato}
{Gallardo} P.~A.,  et al., 2018, in Millimeter, Submillimeter, and Far-Infrared
  Detectors and Instrumentation for Astronomy IX. p. 107082L

\bibitem[\protect\citeauthoryear{{Hand}, {Leauthaud}, {Das}, {Sherwin},
  {Addison}, {Bond}, {Calabrese}, {Charbonnier}, {Devlin}, {Dunkley}, {Erben}
  \& {Hajian}}{{Hand} et al.}{2015}]{lensing}
{Hand} N.,  et al., 2015, \prd, 91, 062001

\bibitem[\protect\citeauthoryear{{Hilton}, {Hasselfield}, {Sif{\'o}n},
  {Battaglia}, {Aiola}, {Bharadwaj} \& {Bond}}{{Hilton} et al.}{2018}]{sz}
{Hilton} M.,  et al., 2018, The Astrophysical Journal Supplement Series, 235,
  20

\bibitem[\protect\citeauthoryear{{Lockman}}{{Lockman}}{2002}]{Lockman}
{Lockman} F.~J.,  2002, in {Stanimirovic} S.,  et al., eds.,  Astronomical
  Society of the Pacific Conference Series Vol. 278, Single-Dish Radio
  Astronomy: Techniques and Applications. pp 397--411

\bibitem[\protect\citeauthoryear{{Louis}, {Grace}, {Hasselfield}, {Lungu},
  {Maurin}, {Addison}, {Ade}, {Aiola} \& {Allison}}{{Louis} et
  al.}{2017}]{Louis}
{Louis} T.,  et al., 2017, \jcap, 6, 031

\bibitem[\protect\citeauthoryear{{Madhavacheril}, {Sehgal}, {Allison},
  {Battaglia}, {Bond} \& {Calabrese}}{{Madhavacheril} et al.}{2015}]{lensing2}
{Madhavacheril} M.,  et al., 2015, \prl, 114, 151302

\bibitem[\protect\citeauthoryear{{Page}, {Jackson}, {Barnes}, {Bennett},
  {Halpern}, {Hinshaw}, {Jarosik}, {Kogut}, {Limon}, {Meyer}, {Spergel},
  {Tucker}, {Wilkinson}, {Wollack} \& {Wright}}{{Page} et al.}{2003}]{Page}
{Page} L.,  et al., 2003, \apj, 585, 566

\bibitem[\protect\citeauthoryear{{Sherwin}, {van Engelen}, {Sehgal},
  {Madhavacheril}, {Addison}, {Aiola}, {Allison}, {Battaglia}, {Becker},
  {Beall}, {Bond} \& {Calabrese}}{{Sherwin} et al.}{2017}]{neutrino}
{Sherwin} B.~D.,  et al., 2017, \prd, 95, 123529

\bibitem[\protect\citeauthoryear{{Swetz}, {Ade}, {Amiri}, {Appel} \&
  {Battistelli}}{{Swetz} et al.}{2011}]{swetz}
{Swetz} D.~S.,  et al., 2011, \apjs, 194, 41

\bibitem[\protect\citeauthoryear{{Tauber}, {Norgaard-Nielsen}, {Ade}, {Amiri
  Parian}, {Banos}, {Bersanelli}, {Burigana}, {Chamballu}, {de Chambure},
  {Christensen}, {Corre}, {Cozzani} \& {Crill}}{{Tauber} et al.}{2010}]{Tauber}
{Tauber} J.~A.,  et al., 2010, \aap, 520, A2

\bibitem[\protect\citeauthoryear{{Thornton}, {Ade}, {Aiola}, {Angil{\`e}},
  {Amiri}, {Beall}, {Becker}, {Cho}, {Choi}, {Corlies}, {Coughlin}, {Datta},
  {Devlin}, {Dicker} \& {D{\"u}nner}}{{Thornton} et al.}{2016}]{thornton}
{Thornton} R.~J.,  et al., 2016, \apjs, 227, 21

\end{thebibliography}

   \end{document}